\documentclass[
  ,draft            % use draft while you are working on the paper
  ,numberedheadings % uncomment this option for numbered sections
  ]
  {aipproc}

\layoutstyle{6x9}

\def\eps{\mbox{$\varepsilon$}}

\def\mtrx #1 #2 #3 #4 {
   \left[ \begin{array}{lr}
           #1 & #2 \\*[1ex]
           #3 & #4
   \end{array} \right] }
\def\mtrxx #1 #2 #3 #4 #5 #6 #7 #8 #9 {
      \left[ \begin{array}{ccc}
            #1 & #2 & #3 \\*[1ex]
            #4 & #5 & #6 \\*[1ex]
            #7 & #8 & #9
      \end{array} \right] }

\let\dsp\displaystyle
\let\eps\epsilon

\let\la\lambda

\let\eps\epsilon

\newcommand{\beq}[1]{  \begin{equation} \label{#1} }
\newcommand{\eeq}{     \end{equation}}

\def\bd#1{\mbox{\boldmath$\displaystyle\mathbf{#1}$} }
\usepackage{amssymb,amsmath}
\usepackage{showlabels}%[inner]
\begin{document}

\title{ Nonlinear evolution equations for degenerate transverse waves in anisotropic elastic solids~}

\classification{\texttt{43.25}}

% \classification{<Replace this text with PACS
%numbers; choose from
%  this list:
%  \texttt{ 43.25 }
  %              \texttt{http://www.aip.org/pacs/index.html}>}

\author{W{\l}odzimierz Doma\'nski}{
  address={ Polish Academy of Sciences, Institute of Fundamental
Technological Research,  \'Swi\c{e}tokrzyska 21, 00-049 Warsaw,
Poland} }

\author{Andrew N. Norris}{
  address={Rutgers University,
Department of Mechanical and Aerospace Engineering, 98 Brett Road,
Piscataway, NJ  08854-8058} }

%\author{<author3>}{
%  address={<common address for author2 and author3>}
%  ,altaddress={<author1 address>} % additional visiting address
%}

\date{}

\begin{abstract}

Transverse elastic waves behave differently in nonlinear isotropic
and anisotropic media. While in the former the quadratically
nonlinear coupling in the evolution equations for wave  amplitudes
is not possible, such a coupling may occur for certain directions in
 anisotropic materials. We identify the expression responsible for
the coupling and we derive coupled canonical evolution equations for
transverse wave  amplitudes in the case of the two-fold and
three-fold symmetry acoustic axes. We illustrate our considerations
by examples for a cubic crystal.

% This template file shows how to use the \texttt{aipproc} class to
% produce a paper with the correct layout for \emph{%
%   AIP Conference Proceedings  6in   x 9in single column}.

% A full description of the features supported by the \texttt{aipproc}
% class can be found in the \texttt{aipguide.pdf} document accompanying
% the distribution.

% Frequently asked questions can be found in the \texttt{FAQ.txt}
% document.
\end{abstract}

\maketitle

%%%%%%%%%%%%%%%%%%%%%%%%%%%%%%%%%%%%%%%%%%%%
%% MAINMATTER
%%%%%%%%%%%%%%%%%%%%%%%%%%%%%%%%%%%%%%%%%%%%

%\section{<A section>}

\section{Introduction}

It is known that transverse elastic waves cannot interact with each
other on the quadratically nonlinear level in an isotropic material
\cite{Gol60}. On the other hand it is also known that in anisotropic
materials this is not the case \cite{Domanski2000}. There are
certain directions in crystals for which quadratically nonlinear
interactions do occur between (quasi-) shear elastic waves. This
manifests itself by the presence of couplings in the evolution
equations for pairs of wave  amplitudes. We show that the two-fold
and three-fold symmetry acoustic axes are examples of these special directions 
along which such a coupling takes place in crystals. We identify the
quantities which when nonzero are responsible for the occurrence of
the coupling between nonlinear elastic shear waves. Then we use the
perturbation method of weakly nonlinear geometric acoustics and we
derive coupled canonical evolution equations for shear wave 
amplitudes of plane waves propagating along  two-fold and
three-fold symmetry acoustic axes.  We illustrate our considerations
by  examples for  cubic crystals.

\section{Plane wave  equations}

Let us consider the system of equations describing  plane waves
propagating in $ \mathbf{n} $ direction in an arbitrary nonlinear
anisotropic hyperelastic material in which both geometrical and
physical nonlinearities are taken into account. This system of
equations can be written as follows \cite{Domanski2000,
Domanski2000b, Domanski2008, Do-No}:
\begin{equation}  \label{quasiplane}
\mathbf{w}_{,\, t} +
\mathbf{A}(\mathbf{w},\mathbf{n})\mathbf{w}_{,\, x} = \mathbf{0},
\end{equation}
with
\begin{equation}
\mathbf{w} = \left[ \begin{array}{l} \mathbf{v}(x,t) \\
\mathbf{m}(x,t)\end{array} \right],\quad
\mathbf{A}(\mathbf{w},\mathbf{n}) = - \left(
\begin{array} {cc}
 \mathbf{0} &  \;\;\;\;\;\; {\frac{1}{\varrho_0}{\ \mathbf{B}(\mathbf{m},\mathbf{n})}} \\
 {\mathbf{I}} & \mathbf{0}
 \end{array}
 \right),\label{matrixA}
\end{equation}
where \begin{equation} {\mathbf B} = \Lambda + {\Psi}{\mathbf m} +
\cdots\, .
% \frac12 \Pi {\mathbf m}{\mathbf m} + \cdots\, .
\end{equation}
%\begin{equation}  \label{e2}
%\dt{\mb{w}} + \mb{A}(\mb{m},\mb{k})\,
%\frac{\partial{\mb{w}}}{\partial {x}} = \mathbf{0},
%\end{equation}
%where \bes
%        \mb{w} = \( \mb{v}(x,t) \\ \mb{m}(x,t) \) \com{and}
%        \mb{A}(\mb{m},\mb{k}) = - \mtrx \mathbf{0} {\ \mb{B}(\mb{m},\mb{k})}
%                {\mb{I}} \mathbf{0},
%\label{6x6mat}\ees
Here  $\varrho_0$ is a constant density\footnote{which, in what
follows, we assume to be equal to 1 for simplicity.}, $\mathbf{v}$
is a velocity, $\mathbf{m}$ is a 1-D strain, $\mathbf{n}$ is the
direction of the plane wave propagation, $\mathbf{I}$ is a $3 \times
3$ identity matrix and $\mathbf{B}$ is a $ 3 \times 3$ symmetric
matrix, which we restrict to consist of zero and first order terms
in strains $\bf{\mathrm{\Lambda}}$ and $\mathrm{\Psi}$,
respectively.

%We demonstrate that the symmetric matrix $\mathbf{B}$ which
%determines plane wave system in any hyperelastic crystal depends
%only on the second and the first derivative of the strain energy in
%3-D. Thus once the strain energy is specified the system of plane
%waves can be written immediately regardless of the anisotropy of the
%material.
Alternatively, in components,
\begin{equation}
 B_{ac}  =c_{abcd}n_bn_d
 + c_{abcdef}  n_bn_dn_f \,m_e + \cdots\, .
 %\frac12
 % N_{abcdefgh}n_bn_dn_fn_h\, m_em_g
%+ \cdots\, .
\end{equation}
where $c_{abcd}$ and  $c_{abcdef}$ are second and third order
elastic constants and $n_a$ and  $m_a$ are components of the vectors
$\mathbf{n}$ and $\mathbf{m}$, respectively. By the standard
reduction of indices we represent the second order constants
with two indices and third order constants with three indices.

We assume that the Christoffel tensor has the form \beq{231}
{\Lambda} = \alpha_1 {\bd k}_1\otimes{\bd k}_1 +
 \alpha_2 {\bd k}_2\otimes{\bd k}_2 +
  \alpha_3 {\bd k}_3\otimes{\bd k}_3 ,
  \eeq
where $\alpha_j >0$ and $\{ {\bd k}_1,{\bd k}_2,{\bd k}_3\}$ is an
orthonormal triad of vectors. The six eigenvalues of ${\bd A}(0)$
are the characteristic speeds of a pair of longitudinal and two
pairs of shear waves.

\begin{align}
\lambda_1 &= - \sqrt{\alpha_1} = - \lambda_2,
\\
\lambda_3 &= - \sqrt{\alpha_2} = - \lambda_4,
\\  \lambda_5 &= -
\sqrt{\alpha_3} = - \lambda_6.
\end{align}
and the right and left  eigenvectors are, respectively,
%\end{document}
\begin{align}\label{270}
 & {\bd r}_{2j-1}  =   \begin{pmatrix}   - \lambda_{2j-1} {\bd k}_j \\    {\bd k}_j  \end{pmatrix},
   \qquad
{\bd r}_{2j}  =  \begin{pmatrix}  - \lambda_{2j} {\bd k}_j \\   {\bd
k}_j    \end{pmatrix} , &
   \nonumber    \\    & &  j=1,2,3,    \\
 & {\bd l}_{2j-1}  = \frac12  \begin{pmatrix}  - \lambda_{2j-1}^{-1} {\bd k}_j \\    {\bd k}_j
   \end{pmatrix} ,    \qquad
{\bd l}_{2j}  = \frac12  \begin{pmatrix}  - \lambda_{2j}^{-1} {\bd
k}_j \\    {\bd k}_j
   \end{pmatrix} . & \nonumber
 \end{align}
 We have ${\bd l}_{i} \cdot {\bd r}_{j} = \delta_{ij}$.
We are interested in cases when the matrix ${\bd A}(0)$   has
coincident eigenvalues corresponding to the speeds of shear waves.
Such waves will be called {\it degenerate}. Hence suppose that $
\alpha_2 = \alpha_3$, so $\lambda_s = \lambda_{s + 2}$ for $s =
3,4$. In such a case $\bd n$ is an {\it acoustic axis} (see
\cite{khatkevich1962}, or \cite{Norris04b}).

\section{Coupled evolution equations}
Let us consider the initial-value problem for the system of plane
waves elastodynamics (\ref{quasiplane}), (\ref{matrixA}) with
perturbed initial data which are  of compact support ($\epsilon$ is
a small parameter):
\begin{equation}\label{Cauchy-eps}
\left\{
\begin{array}{l}
\mathbf{w}_{,\, t}^{\eps} +
\mathbf{A}(\mathbf{w}^{\eps},\mathbf{n})\mathbf{w}^{\eps}_{,\, x} =
\mathbf{0}, \\[2ex]
 %\partial_t {\mathbf w}^{\eps} + \mb{A}(\mb{w}^{\eps}, \mathbf k) \partial_x {\mathbf w}^{\eps} = 0, \\[2ex]
{\mathbf w}^{\eps}\,_{|\,t = 0}  =  \eps\,\mathbf{w}_{1}(x,x/\eps).
\end{array}
\right.
\end{equation}
We apply the asymptotic method of weakly nonlinear geometric
acoustics. The method combines the use of a multiscale perturbation
with a small amplitude, high frequency asymptotics
\cite{Domanski2000}. Assume that the asymptotic expansion is of the
form \beq{exp} {\bd w}(t,x) = \epsilon \bigg( \sigma_s(t,x,\eta)
{\bd r}_s +\sigma_{s+2}(t,x,\eta) {\bd r}_{s+2}\bigg) +O(
\epsilon^2), \qquad s=3,4, \eeq with $\theta =
\epsilon^{-1}(x-\lambda t)$,  $ \lambda = \lambda_s =
\lambda_{s+2}$, and where $\sigma_s, \sigma_{s + 2}$ are the unknown
amplitudes. Inserting (\ref{exp}) into (\ref{Cauchy-eps}) and
applying the method of multiple-scale asymptotics (see \cite{Do-No}), we
obtain the following coupled evolution equations for the amplitudes
$\sigma_s$ and $\sigma_{s + 2}$:
\begin{equation}\label{general}
\left\{
\begin{array}{rrl}
\dsp (\sigma _{s})_{,t} + \lambda_s(\sigma _{s})_{,x} + \dsp
\Gamma_{s} \sigma _{s} (\sigma _{s})_{,\theta } +
\Gamma^{s+2}_{s}(\sigma _{s} {\sigma }_{s+2})_{,\theta }
 + \Gamma^{s}_{s+2} \sigma _{s+2} (\sigma _{s+2})_{,\theta} =0, &  &  \\[1ex]
\dsp ({\sigma _{s+2}})_{,t } +  \lambda_s(\sigma _{s})_{,x}  +  \dsp
\Gamma^{s+2}_{s} \sigma _{s} (\sigma _{s})_{,\theta } +
\Gamma^{s}_{s+2}(\sigma _{s} {\sigma }_{s+2})_{,\theta }
 + \Gamma_{s+2} \sigma _{s+2} (\sigma _{s+2})_{,\theta} =0.
\end{array}
\right.
\end{equation}
Here we denote  $\Gamma^{s}_{s,s} \equiv  \Gamma_{s}$, and we use
the relations which hold for elastodynamics

\begin{equation}
\Gamma^{s+2}_{s}  \equiv  \Gamma^{s}_{s,s+2} = \Gamma^{s+2}_{s,s},
\;\;\;\;\;\; \;\; \Gamma^{s}_{s+2} \equiv \Gamma^{s+2}_{s,s+2} =
\Gamma^{s}_{s+2,s+2},
\end{equation}
%\begin{align}\label{54}
%\Gamma^{s+2}_{s}  \equiv  \Gamma^{s}_{s,s+2} &= \Gamma^{s+2}_{s,s}
%\\
%\Gamma^{s}_{s+2} \equiv \Gamma^{s+2}_{s,s+2} &=
%\Gamma^{s}_{s+2,s+2}, \label{55}
%\end{align}
where  the interaction coefficients are defined
\beq{44} \Gamma^j_{p,q}  = {\bd
l}_j\cdot \big( \nabla_{{\bd w}} {\bd A}({\bd w}) {\bd r}_p{\bd r}_q
\big)|_{{\bd w} = {\bd 0}}. \eeq
\subsection{Two-fold symmetry acoustic axis}
In the case of a two-fold symmetry acoustic axis, the evolution
equations further simplify and it turns out (see \cite{Do-No}) that
only two coefficients $ \Gamma_{s} $  and $\Gamma^{s}_{s+2} $
characterize the nonlinear terms in the coupled evolution equations
(\ref{general}) for a pair of amplitudes of (quasi-)shear waves, in
that case.
\subsection{Three-fold symmetry acoustic axis}
In the case of a three-fold symmetry acoustic axis only one
coefficient  $\Gamma_{s}$ characterizes the nonlinear terms in the
evolution equations for a pair of amplitudes of plane degenerate
shear waves, and the coupled evolution equations are  as follows:
\begin{equation}\label{cBurgers1}
\left\{
\begin{array}{rrl}

\dsp ({\sigma _{s}})_{,t } +  \lambda_s(\sigma _{s})_{,x} + \dsp
\Gamma _{s}\left[\sigma _{s} (\sigma _{s})_{,\theta } - \sigma
_{s+2} (\sigma _{s+2})_{,\theta}\right] =0, &  &  \\[1ex]
\dsp (\sigma _{s + 2})_{,t} +  \lambda_s(\sigma _{s + 2})_{,x} -\dsp
\Gamma _{s}(\sigma _{s}\,{\sigma }%
_{s+2})_{,\theta }=0.

%\dsp ({\sigma _{s}})_{,t } +  \lambda_s(\sigma _{s})_{,x}  - \Gamma _{s + 2}(\sigma _{s}\,{\sigma }%
%_{s+2})_{,\theta }=0, &  &  \\[1ex]
%\dsp (\sigma _{s + 2})_{,t} +  \lambda_s(\sigma _{s + 2})_{,x} -\dsp
%\Gamma _{s + 2}\left[\sigma _{s} (\sigma _{s})_{,\theta } - \sigma
%_{s+2} (\sigma _{s+2})_{,\theta}\right] =0.
\end{array}
\right.
\end{equation}

%%%%%%%%%%%%%%%%%%%%%%%%%%

\section{Examples}

In the examples below we consider a cubic crystal of class
\emph{m3m} in which the strain energy $W$ is characterized by three
second order and six third order elastic constants (see
\cite{Domanski2000, Domanski2000b, Domanski2008, Do-No})
$$
W = W(c_{11}, c_{12}, c_{44}, c_{111},  c_{112}, c_{144}, c_{123},
c_{166}, c_{456}).
$$

{\bf Example 1}.
%\subsubsection{Example 1}
Here we present the case where the propagation direction of the
plane wave is a two-fold symmetry axis ${\mathbf{n}} = \dsp
\tfrac{1}{\sqrt{2}}[1\, 1\, 0]$ which is not an acoustic axis. The speeds of shear waves are (see \cite{Domanski2000, Domanski2000b, Domanski2008, Do-No})
$$
%\begin{array}{rcl}
\la_3 = -\sqrt{\dsp\frac{c_{11} - c_{12}}{2}} = -\la_4, \;\;\;\;\;\;
\la_5 = -\sqrt{c_{44}} = -\la_6,
%\end{array}
$$
hence $\la_3 \neq \la_4 \neq \la_5 \neq \la_6$, and $\Gamma_s = 0$
for $s = 3, 4, 5, 6.$ It turns out that there is no coupling and
moreover one can show that decoupled evolution equations with cubic
nonlinearity describe the propagation of shear waves in this case
(see \cite{Domanski2000}).

{\bf Example 2}.  Consider now the case of a three-fold symmetry
axis
 ${\mathbf{n}} = \dsp \tfrac{1}{\sqrt{3}}[1\,
1\, 1]$ which is the acoustic axis. In this case we obtain (see
\cite{Domanski2000, Domanski2000b, Domanski2008, Do-No}) coupled
equations (\ref{cBurgers1}) with
$$
\begin{array}{rcl}
\la_3 = \la_5 &=& \dsp -\sqrt{\frac{c_{11} -  c_{12} +
 c_{44}}{3}}= -\la_4= -\la_6.
\end{array}
$$

$$ \Gamma_3 =  - \dsp \frac{1}{18 \sqrt{2}}\left[ \frac{c_{111} + 2
c_{123} - 2 c_{456} - 3 (c_{112} -  c_{144} +
c_{166})}{\la_3}\right] = - \Gamma_4.
$$

%%%%%%%%%%%%%%%%%%%%%%%%%%%%%%%%%%%%%%%%%%%%%%%%

\begin{theacknowledgments}W.D. was partially supported by the Polish State Committee grant No. 0 T00A
014 29.
  %Infandum, regina, iubes renovare dolorem, Troianas ut opes et
 % lamentabile regnum cruerint Danai; quaeque ipse miserrima vidi, et
 % quorum pars magna fui. Quis talia fando Myrmidonum Dolopumve aut duri
%  miles Ulixi temperet a lacrimis?
\end{theacknowledgments}


\begin{thebibliography}{9}


\bibitem[Domanski(2000)]{Domanski2000}
W.~Doma\'nski,
%\newblock Weakly nonlinear elastic plane waves in a cubic crystal.
\newblock \emph{Contemp. Math.} {\bf  255}, 45--61 (2000).

\bibitem[Domanski(2000b)]{Domanski2000b}
W.~Doma\'nski,   ``Propagation and interaction of finite amplitude
elastic waves in a cubic crystal,''    in \emph{ Nonlinear Acoustics
at the Turn of the Millenium}, edited by W.~Lauterborn and T.~Kurz,
AIP Conference Proceedings 524, American Institute of Physics,  New
York, 2000, pp. 249--252.

\bibitem[Domanski(2008)]{Domanski2008}
W.~Doma\'nski, \newblock \emph{Wave Motion} {\bf  45}, 337--349
(2008).

\bibitem[Domanski-Norris(2008)]{Do-No}
W.~Doma\'nski, and A.~N. Norris, to appear 2008.

\bibitem[Goldberg(1960)]{Gol60}
Z.~A.~Goldberg,
%\newblock \emph{On selfinteraction of plane longitudinal and transverse waves}.
\newblock \emph{Soviet. Phys. Acoust.} {\bf  6},  307--310 (1960) in Russian.

\bibitem[Fedorov(1968)]{fed68}
F.~I. Fedorov,
\newblock \emph{Theory of Elastic Waves in Crystals}.
\newblock Plenum Press, New York, 1968.


\bibitem[Khatkevich(1962)]{khatkevich1962}
A.~G. Khatkevich,
%\newblock The acoustic axis in crystals.
\newblock \emph{Sov. Phys. Crystallogr.}, {\bf 7},  601--604, (1962).


\bibitem[Norris(1991)]{Norris91}
A.~N. Norris,
%\newblock Symmetry conditions for third order elastic moduli and implications
%  in nonlinear wave theory.
\newblock \emph{J. Elasticity}, {\bf 25}, 247--257 (1991).

\bibitem[Norris(1999]{Norris99} A.~N. Norris, ``Finite amplitude waves in solids,''
  in \emph{ Nonlinear Acoustics}, edited by M.~F.~Hamilton
and D.~T.~Blackstock,  Academic Press, San Diego,  1999, pp.
263--277.

\bibitem[Norris(2004)]{Norris04b}
A.~N. Norris,
%\newblock Acoustic axes in elasticity.
\newblock \emph{Wave Motion}, {\bf 40}, 315--328 (2004).


\end{thebibliography}
\end{document}